\documentclass[12pt]{iopart}

  \expandafter\let\csname equation*\endcsname\relax
  \expandafter\let\csname endequation*\endcsname\relax

\usepackage{graphicx}
\usepackage{iopams} 
\usepackage{amsmath}
\usepackage{braket}
\usepackage{color}
\usepackage{ulem}

\begin{document}

\title{Interaction between localized vegetation patches and gaps in  water-limited environments}

\author{M. Tlidi$^a$, E. Berr\'ios-Caro$^b$, D. Pinto-Ramo$^b$, A.G. Vladimirov$^c$, and  M. Clerc$^b$ }

\address{$^a$D\'epartement de Physique, Facult\'e des Sciences, Universit\'e Libre de Bruxelles (U.L.B.), CP 231, 
Campus Plaine, B-1050 Bruxelles, Belgium}
\address{$^b$Departamento de F\'isica and Millennium Institute for Research in Optics, Facultad de Ciencias F\'isicas y Matem\'aticas,
Universidad de Chile, Casilla 487-3, Santiago, Chile}

\address{$^c$Weierstrass Institute, Mohrenstrasse 39, 10117 Berlin, Germany}

\ead{mtlidi@ulb.ac.be}
\vspace{10pt}
\begin{indented}
\item[]August 2019
\end{indented}

\begin{abstract}
Close to the critical point associated with nascent of bistability
and large wavelength pattern forming regime, {\it the Lifshitz point},
the dynamics of many ecological spatially extended systems can be reduced to a simple partial differential equation. 
This weak gradient approximation is greatly useful for the investigation of localized vegetation patches 
and gaps. In this contribution, we present a general derivation of the most simple vegetation model 
without any specification of the shape of Kernel used to describe the facilitative and the competitive 
interactions between individual plants.  The coefficients of the obtained model depend 
on the choice of the form of the Kernel under consideration. Based on this  simple vegetation model, 
we focus more on gaps and patches 
interaction. In the case of gaps, the interaction alternates between attractive and repulsive depending on the distance
separating the gaps. This allows for the stabilization of bounded states and clusters of gaps. 
However, in the case of localized patches, the interaction is always repulsive. 
In this former case, bounded
states of patches are excluded. The analytical formula of the interaction potential is derived and reviewed for both types of interactions and checked by numerical investigation of the model equation.
\end{abstract}

\section{Introduction}
Spatial fragmentation of landscapes is an inherent characteristic of semi- and arid-ecosystems. 
In these regions, vegetation populations exhibiting non-random two-phase structures where 
high biomass density regions are separated by sparsely covered or even bare ground. 
They cover extensive arid- and semi-arid areas worldwide \cite{Meron2015,Deblauwe_2008}. 
The most common are spotted patterns, more or less circular patches, the second consists of 
bands or arcs (or even spirals), and the third are gaps that reside in spots of bare soil. 
The climate of these regions is characterized by water scarcity typical the annual rainfall lies between 50 to 750 mm. 
Aridity refers not only to water-limited resources but can be originated from nutrient-poor territories.  
These self-organized structures consist of spatially periodic distributions of patches, 
stripes or gaps that occupy the whole space available in a landscape.  
Transition sequence between this vegetation pattern has been established.  
As the level of the aridity level is increased, the first pattern that appears consist of 
a spatial periodic distribution of gaps followed by stripes (or labyrinth) and spots. 
This generic scenario has been predicted from various ecological models \cite{LEJEUNE99,vonHardenberg2001,Gilad2004,LEJEUNE04,Rietkerk,Gowda14,Getzin,Mander17,Yizhaq}.  
It is now generally admitted that this adaptation to hydric stress involves a symmetry-breaking 
instability leading to the establishment of a stable spatial periodic pattern \cite{Lefever1997},
 that is, biomass has to self-organize to optimize the use of the scarce resources.

Vegetation patterns are not always periodic; they can be aperiodic and localized in
space. They consist of either localized patches of vegetation, randomly distributed on
bare soil  \cite{Lejeune2002,Ehud2004,Meron07,Couteron_14,Escaff_2015} or, on the contrary, consist of 
localized spots of bare soil,
randomly distributed in an otherwise uniform vegetation cover \cite{TLV08,Tarnita,Ruiz-Reynes}. An example of such botanical 
self-organization phenomenon is shown in Fig. 1. 
This figure illustrated the phenomenon of self-organization is generic, 
as it occurs on different spatial scales, kind of soil, continents,  and for different type of vegetations.
Spatial localized structures or patterns are better known 
in contexts of a physicochemical rather than biological system (see overviews \cite{Murray,TlidiIthme_2007,Knobloch07,Knobloch08,Akhmediev2008,Ridolfi2011,descalzi,Tlidi2004,Tlidiclerc,Chembo,Prigo1}

\begin{figure}[h]
\begin{center}
\includegraphics[width=0.67\textwidth]{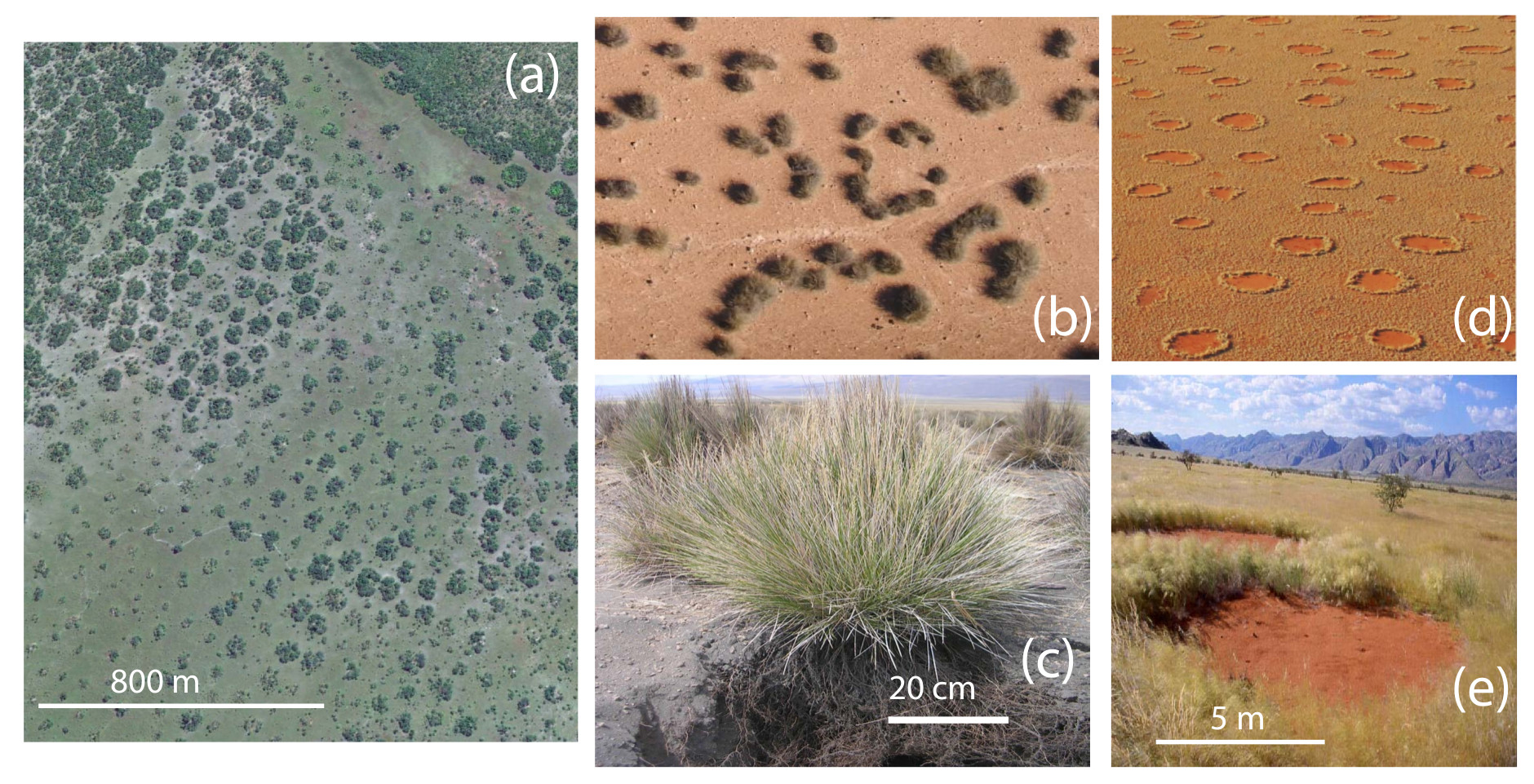}
\end{center}
\caption{Localized vegetation patches and gaps. Images of self-organization of localized patches
obtained using Google Earth Pro (a,b).
(a) Zambia, Southern Africa (13$^{\circ}$46'49.07'' S, 25$^{\circ}$16'56.97'' E). 
(b) Ivory Coast, West Africa (7$^{\circ}$14'53.01'' N, 6$^{\circ}$06'27.83''W).
(c)  Festuca orthophylla (observed in the Sajama National Park, Bolivia
(d,e) Pro-Namibia zone of the west coast of Southern Africa (courtesy of N. Juergen). 
Random distribution Gaps or fairy circles 
(e) two interacting gaps (photography: courtesy of M. Johnny Vergeer)}.
\label{pictures}
\end{figure}
In what follows, we consider the propagation-inhibition type of models under inhomogeneous and isotropic environmental conditions. The effect of the slope of the ground, water dynamics in surface or depth, the wind, or the course of
the sun is not considered and neglected in this approach. The model allows for the genesis of patterns based solely on the
intrinsic dynamics of the vegetation. In other words, the wavelength of the pattern that emerges from the symmetry-breaking instability depends solely on the
aridity and on  facilitative and competitive interactions ratio.
In the first part of this paper, we present a general derivation of a real partial differential equation for the vegetation without any specification of the type of the Kernel function associated with the facilitative and the competitive interactions. We show that this model does not depend on the kind of Kernel providing that its Taylor series converge. Typically, we use Gaussian and exponential type of Kernels. In the second part, we focus on the interaction between patches and gaps. Analytical computations allow for the construction of interaction potential between two well-separated localized patches and gaps.

The paper is organized as follows. After an introduction, we present the propagation-inhibition model in Sec.~2. 
We derive a real partial differential equation in Sec.~3. 
The interaction between two well-separated isolated patches and gaps is performed in Sec.~4 
and Sec.~5, respectively. We conclude in Sec.~6.

\section{Space-time dynamics of vegetation in scarce environments}
Physicochemical systems models are derived from first principles. However, the absence of the first principles for biological systems in general, and ecological ecosystems render mathematical modeling complex.  Most of the mathematical models proposed in the literature are models based on water transport \cite{Mauchamp,Thiery,Dunkerley,Klausmeier,Okayasu,Sherratt2005}.
In contrast, the theory introduced by Lefever and coworkers is grounded on the balance between facilitation and competition interactions exerted by plants themselves, through their above-ground and below-ground parts \cite{Lefever1997}.  Homogeneous and isotropic environmental conditions are assumed. Besides, no water dynamics neither in surface nor in depth is assumed. 
This theoretical approach is a generalization of the paradigmatic logistic equation \cite{Verhulst,Mawhin1,Mawhin2} with non-local facilitation, competition and seed dispersion \cite{Lefever1997,TLV08}. This model predicts the fragmentation of a uniformly cover when the radius of the root system is larger than the canopy radius \cite{Lefever1997}. Indeed, when the level aridity is increased, superficial roots track scarce water far beyond the limits of aerial parts of the plant. The wavelength associated with this symmetry breaking instability is intrinsic and depends only on the structural parameters such as canopy-to-rhizosphere radius ratio. For trees and shrubs, data from arid regions show that the canopy-to-rhizosphere radius ratio may be as small as 1/10. Besides front propagation leading to the stabilization of localized vegetation patches and gaps have been reported in  \cite{Fernandez-Oto13,Martinez-Garcia13,Martinez-Garcia14,Fernandez-Oto14,Colet14,LENDERT14,Dornelas}.

The existence of such botanical self-organization is not restricted to a special kind of plant. They may entirely consist of grass, shrubs or trees (cf. Fig.~1). They are also not specific to a special kind of soil, which can go from sandy to silty or clayey. To simplify further the modeling of ecosystems, we consider that all plants are mature and we neglect age
classes since individual plants grow on much faster time scale comparing to the time scale of the formation of
periodic vegetation pattern. The only variable is the vegetation biomass density
which is defined at the plant level. The spatiotemporal 
evolution of the normalized biomass $b(\bf{r},t)$ obeys to the following integrodifferential equation 
\begin{equation}
\partial_t b=B_1-B_2 +Dispersion.\label{TW1}
\end{equation}
Time has been scaled such that the characteristic time of the growth process is unity.
The term  $B_1$ models the rate at which the biomass increases and saturates.  
The biomass losses which $B_2$ describes death or destruction by grazing, fire, 
termites, or herbivores. The last term describes seed dispersion. 
\begin{equation} \label{dispersion}
Dispersion= D \int \left[ \phi_{in}b(|{\bf{r}}+{\bf{r'}}|,t)-\phi_{out}b(|{\bf{r'}}|,t) \right] d{\bf{r'}},
\end{equation}
$D$ is the rate at which plants diffuse,  $\Phi_{in}$ and $\Phi_{out}$ are the dispersion kernels weighting the incoming and outgoing seed fluxes between neighboring plants according to their distance $|{{\bf r^{\prime}}}|$. 
The plant-to-plant
feedback interactions can be written as:
\begin{equation} \label{Kernelsfc}
B_1=b\left(1-b\right)\mathcal{M}_f,  { \mbox{ and  }}  B_2=-\mu b\mathcal{M}_c
\end{equation}
The biomass gains  $B_1$ account for the biomass productions via  dissemination, germination 
and other natural mechanisms that tend to increase the biomass.  This exponential growth is 
impossible to maintain over long time evolution for any population including non cognitive 
populations such as plants because resource scarcity.  The logistic term  proportional 
to $(1-b)$ impedes the biomass production that describes the fact that the biomass 
cannot exceed the carrying capacity. The biomass losses which $B_2$ 
describes death or destruction by grazing, fire, termites, or herbivores. 
The last term describes seed dispersion.  The plat-to-plant nonlocal interactions are
\begin{equation}
M_{f,c}({\bf{r}},t) = \exp{\left(\chi_{f,c} \int \phi_{f,c} b(|{\bf{r}} + {\bf{r'}}|,L_{f,c})  d{\bf{r'}}  \right)},
\end{equation}
where kernels are normalization such as $\int \phi_{f,c}  d{\bf{r'}} =1$.  The parameters 
${\chi_{c,f}}$ are the interaction strength associated with the competitive and facilitative, respectively. The parameter $\mu$ measures the resources scarcity often called the aridity parameter. The nonlocal $M_{f}({\bf{r}},t)$ function describes interactions facilitating the growth of the biomass such as seed production, germination and other mechanisms that facilitate the increase of biomass density. This positive feedback operates oven a distance  $L_f$ of the order of the plant's aerial structures (the radius of the crown or the canopy) involving, in particular, a reciprocal sheltering of neighboring plants
against arid climatic conditions. The  nonlocal $M_{c}({\bf{r}},t)$ models the plant-to-plant competitive interactions that on the contrary tends to enhance biomass decay. This negative feedback operates operate over distances of the order of the size of roots, i.e., the rhizosphere radius $L_c$. Plants compete through their roots for resources. In other words, the rhizosphere activity of individual plant
tends to cut out its neighbors from resources.

The kernel function that characterizes the nonlocal facilitative and competitive interactions $\phi_{f,c}$ 
and seed dispersion $\phi_{in,out}$ must satisfy three general conditions: 
(i) the Kernels must be equal to one when the zero biomass limit, i.e., $\lim_{b \to 0}\phi_{f,c,in,out} = 1$, (ii) they should be normalized to ensure that that the homogeneous steady states of Eq. (1) 
are independent of the range of interactions of Kernels, i.e.,  $\int  \phi_{f,c,in,out}(|{\bf{r'}}|)  d{\bf{r'}}=1$, 
and (iii) we assume that the Kernel are symmetric and  even functions, i.e., $\phi_{f,c,in,out}(|{\bf{r'}}|)= \phi_{f,c,in,out}(-|{\bf{r'}}|)$. A this stage, the spatial shape the Kernel  $\phi_{f,c,in,out}(|{\bf{r'}}|)$ has not been yet specify. Generally speaking, Kernels or influence functions can be classify into two types  either weak and strong. If the kernel function
decays asymptotically to infinity more slowly (faster) than
an exponential function, the nonlocal coupling is said to be
strong (weak) \cite{Escaff_2011,Escaff_2015}. Let rewrite Eq. (1) as
\begin{equation}
\partial_t b= b\left(1-b\right)\mathcal{M}_f-\mu b\mathcal{M}_c+D \int \left[ \phi_{in}b(|{\bf{r}}+{\bf{r'}}|,t)-\phi_{out}b(|{\bf{r'}}|,t) \right] d{\bf{r'}}. 
\label{INTEGRO}
\end{equation}
The homogeneous steady-state solutions of Eq. (1) are given by 
\begin{equation} \label{HSS_1}
\mu=(1-b_s) \exp{(\Lambda b_s)},
\end{equation}
where $\Lambda=\chi_{f}-\chi_{c}$ is the feedback balance often called the cooperativity parameter.

In the next section, we perform the nonlinear analysis of the  spatiotemporal 
evolution of the biomass described by the logistic equation with nonlocal interaction 
between plants (Eq. 1) in the double limit of bistability and close to
the vegetation pattern forming threshold.
This type of critical point in the parameter space is known as {\it Lifshitz point} \cite{Hornreich,Cross93}, 
which was initially proposed in the context of pattern formation in magnetic systems \cite{Hornreich}.

\section{A nonvariational Swift-Hohenberg for vegetation pattern formation}
The purpose of this section is to present the derivation of
the non-variational Swift-Hohenberg \cite{LEJEUNE99}  from the integrodifferential model for 
the biomass Eq.~(\ref{INTEGRO}). 
In the case of a Gaussian Kernel, a detailed derivation of this generic local model 
equation is presented in \cite{TLV08}. Another nonvariational Swift-Hohenberg equation has been derived for spatially extended systems \cite{Kozyreff2003,Clerc2005,Durniak,Kozyreff2007,Alvarez-Socorro}. However, a variational Swift-Hohenberg equation has been derived first in the context of fluid mechanics \cite{SH77} and since then it constitutes a paradigmatic model in the study of the periodic or
localized structure and localized patterns. It has been derived for that purpose in
other fields of natural science, such as  electroconvection \cite{Richter2005}, mechanics \cite{Wadee2002,Hunt2003},  chemistry \cite{Hilali}, plant ecology \cite{Lefever09}, and nonlinear
optics \cite{Mandel93,Tlidi93,Tlidi94}.

In what follows, we explore the space-time dynamics in the vicinity of  the Lifshitz
critical point associated with bistability where the homogeneous steady states $b_s$ solution of 
Eq. (\ref{HSS_1}) undergo a second-order critical point marking the onset of a hysteresis loop where the inflection point (with infinite
slope) of the biomass response curve corresponding to
the critical cooperatively for the onset of bistability $ \Lambda_c=1$. At this critical point, the critical value of the biomass is $b_c=0$ and the aridity parameter is $\mu_c=1$.  
We first define a small parameter $\epsilon$ which measure the distance from criticality as
\begin{equation} \label{expan_coop}
\Lambda=\Lambda_c+\lambda_0 \epsilon^{1/2}=1+\lambda_0 \epsilon^{1/2}\cdots.
\end{equation}
We next expand the aridity parameter, as well as the dependent
biomass density as
\begin{equation} \label{expan_aridity}
\mu=1+\lambda_0 \epsilon+\cdots,  { \mbox{ and  }} b({\bf{r}},t) =\epsilon u_0({\bf{r}},t)+\cdots.
\end{equation}

We finally we use the following scaling for the strength of the competitive feedback and the dispersion coefficient as
\begin{equation} \label{expan_strength}
\chi_c=\chi_0+\chi_1 \epsilon^{1/4}+\cdots,  { \mbox{ and  }} D =\epsilon^{3/4}+\cdots.
\end{equation}
Since, we are interested in the regime where the biomass density is small, 
we can then perform a Taylor expansion of the exponentials appearing in Eq.~1 as
\begin{eqnarray} \label{Taylor1}
M_{f,c}({\bf{r}},t) &=& 1+ \chi_{f,c}\int \phi_{f,c} b(|{\bf{r}} + {\bf{r'}}|,L_{f,c})  d{\bf{r'}}\nonumber \\&+&\frac{1}{2} \left( \int \phi_{f,c} b(|{\bf{r}} + {\bf{r'}}|,L_{f,c})  d{\bf{r'}}\right)^2+\cdots.
\end{eqnarray}
In the same manner, we perform a Taylor expansion of the dispersion Kernels  $\phi_{in,out}$.  
The biomass at the position ${\bf{r}} + {\bf{r'}}$ is also expanded as 
\begin{eqnarray}  \label{Taylor2}
 b({\bf{r}} + {\bf{r'}},t)&=&b({\bf{r}},t)+\left[{\bf{r'}}\cdot\nabla +\frac{1}{2}({\bf{r'}}\cdot \nabla)^2+\frac{1}{3!}({\bf{r'}}\cdot\nabla)^3\right]b({\bf{r}},t)\nonumber \\
&+&\frac{1}{4!}({\bf{r'}}\cdot\nabla)^4b({\bf{r}},t)+\cdots.
\end{eqnarray} 
Since the Kernel must be normalized and be even function with respect to the spatial coordinate ${\bf{r}}=(x,y)$, we have
\begin{equation} \label{Taylor3}
\int \phi_{f,c} b(|{\bf{r}} + {\bf{r'}}|,L_{f,c})  d{\bf{r'}}=b({\bf{r}},t)+{\bf C}_2^{f,c} \cdot \nabla^2 b({\bf{r}},t)
+{\bf C}_4^{f,c}  \cdot \nabla^4 b({\bf{r}},t)
+\cdots
\end{equation}
where 
\begin{equation} \label{Taylor3}
{\bf C}_n^{f,c} =\int \phi_{f,c} ({\bf r}') \frac{{\bf{r'}}^n}{n!} d{\bf r}'.
\end{equation}
We seek corrections to the steady states at criticality
that depends on time and space through the slow variables
\begin{equation} \label{scaling}
(\tilde{x}, \tilde{y})=(x,y) \epsilon^{1/2}  { \mbox{ and  }} \tilde{t}=\frac{\epsilon}{2}t 
\end{equation}
and therefore $\nabla^2=\epsilon^{1/2} \tilde{\nabla}^2$ and $\partial_t= \epsilon \partial_{\tilde{t}}/2$. Replacing Eqs. (\ref{Taylor1}-\ref{scaling})  in Eq. (\ref{INTEGRO}), and expand in series of $\epsilon$, to satisfy the solvability condition  at the order $\epsilon^{5/4}$, we obtain 

\begin{equation} \label{leasinf_Order}
\chi_0=\frac{C_2^f}{C_2^c-C_2^f}
\end{equation}
At higher order inhomogeneous problem (order $\epsilon^{3/2}$), we obtain the following partial differential equation
\begin{equation} 
\label{SH}
\partial_t u=-u(\eta-\kappa u+u^2)+(\delta-\gamma u)\nabla^2u-\alpha u\nabla^4u,
\end{equation}
where
\begin{eqnarray} \label{SH1}
\eta&=&2\mu_0, \quad \kappa=2\Lambda_0,  \quad \delta=2pC_2^{in},   \nonumber \\
\gamma&=&2\chi_1(C_2^c-C_2^f),  \quad \alpha =2\left(\frac{C_2^cC_4^f-C_2^fC_4^c}{C_2^c-C_2^f}\right).
\end{eqnarray}

From equation Eq. \ref{SH}, we finally obtain the simple local vegetation model \cite{TLV08}

In Eq. (\ref{SH}),  the parameter $\eta$ and $\kappa$ are, respectively, 
the aridity parameter and the cooperatively, i.e., the feedback balance. 
These two parameters do not depend on the nature and the shape of the 
Kernels used to describe nonlocal facilitative and competitive interactions, $\phi_{f,c}$, 
and the dispersion kernels weighting the incoming and outgoing seed fluxes $\phi_{in,out}$. 
However, the parameters $\delta$, $\gamma$, and $\alpha$ as we will see in the next 
subsections that they strongly depend on the spatial form of the kernel. In the case of strong nonlocal coupling 
mediated by a Lorentzian type of Kernel, the above real order parameter cannot 
leading to the Swift-Hohenberg, this is because of Taylor series do not converge in the case  
of Lorentzian type of Kernel. Namely, the integral cannot be expressed as a gradient expansion. 
Therefore, the analytical expression of the coefficient $\delta$, $\gamma$, and $\alpha$  will be given in the case of 
Gaussian, step, and exponential type of Kernels. 

Equation (\ref{SH}) holds for a Gaussian, exponential, and bounded influence function and 
other types of Kernels provided that theirs Taylor series converges. 
Note that Eq. (\ref{SH}) has been proposed in an early report \cite{LEJEUNE99}, 
and derived by using a multiple scaling analysis \cite{TLV08}. It has been shown that Eq.  (\ref{SH}) support both localized patches  \cite{Lejeune2002} and localized gaps \cite{TLV08}. A single or more localized patches can suffer from a curvature instability leading to the self-replication phenomenon 
\cite{Bordeu_SR,Bordeu_15,Bordeu_chapter16,Bordeu_EO_2018}. Recently, Eq. (\ref{SH}) has been recovered \cite{OTO2019} by using 
the water-limited vegetation model proposed by Meron and collaborators \cite{Meron_2016}.

The derivation of the model Eq. (\ref{SH}) is independent of the shape and the form of the kernel considered. To establish the connection between the physical parameters and the coefficients $\delta$, $\gamma$ and $\alpha$, we choose for simplicity Gaussian Kernels for both facilitation, competition and seed dispersion as
\begin{equation}  \label{Gaussian}
\phi_{f}=\frac{1}{\pi }\exp{(-|{\bf{r'}}|^2)},  \qquad
\phi_{c}=\frac{\epsilon_1}{\pi}\exp{\left(-\epsilon_1|{\bf{r'}}|^2\right)},   
{ \mbox{ and  }} 
\phi_{in}=\frac{\epsilon_2}{\pi}\exp{\left(-\epsilon_2|{\bf{r'}}|^2\right)},
\end{equation}
with $\epsilon_1=L_a^2/L_c^2$ and  $\epsilon_2=L_a^2/L_d^2$ where $L_d$ is the length associated with seed dispersion. For the Gassian Kernel we have
$$
\delta=\frac{p}{2\epsilon_2}, \gamma=\frac{\chi_1(1-\epsilon_1)}{2\epsilon_1},  { \mbox{ and  }} \alpha=\frac{1}{16\epsilon_1}
$$
The coefficients  $\delta$, $\gamma$ and $\alpha$ that appear in Eq.(\ref{SH})  depend on the form and the shape of the Kernels. However, the coefficients  $\eta$ and $\kappa$ do not depend on a particular choice of the Kernels.

\section{Bifurcation diagram and periodic vegetation patterns}
The homogeneous steady states solutions of Eq.  (\ref{SH}) satisfy 
\begin{equation} \label{hhhs}
u_0=0,  { \mbox{ and  }}  u_{\pm}=\frac{\kappa\pm\sqrt{\kappa^2-4\eta}}{2}.
\end{equation}
The spatially uniform bare state $b_0$ corresponds to a state devoid of vegetation. The homogeneously covered vegetation state is denoted by $u_{+}$. These two states are separated by an unstable homogeneous cover with lower biomass $u_{-}$. 

\begin{figure}[h]
\begin{center}
\includegraphics[width=0.67\textwidth]{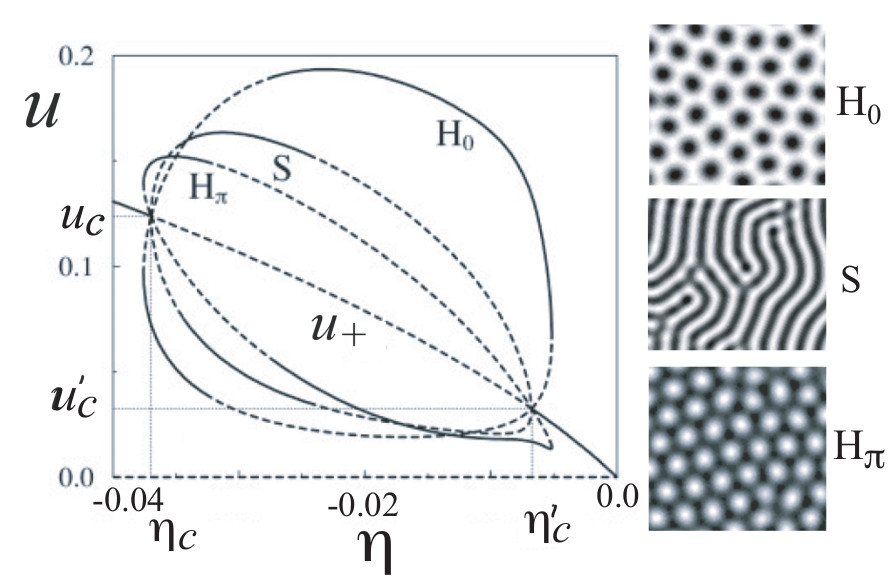}
\end{center}
\caption{Bifurcation diagram of model Eq. (\ref{SH}) obtained for 
the monostable regime ($\eta=-0.175<0$) as a function of the aridity parameter $\eta$. 
Others parameters are $\delta=0.005, \gamma=0.5$, and $\alpha=0.25$. 
The sequence hexagons $H_{\pi}$, Stripes  S and the   hexagons $H_{0}$ are 
obtained for $\eta=-0.035$, $\eta=-0.03$, and $\eta=-0.02$, respectively. 
These two dimensional structures have been obtained by direct numerical simulation of Eq. (\ref{SH}) 
using the finite differences scheme with Runge-Kutta order-4 algorithm and
periodic boundary conditions.  Black color corresponds to the highest biomass density.    }
\label{bifdiam}
\end{figure}

Let us analyze the effect of small amplitude disturbance of a given ecosystem 
around the uniformly  vegetated state $u_{+}$. Mathematically this means that 
we have to apply small amplitude fluctuations $\delta  u({\bf{r}},t)$ around the state   $u_{+}$ of the form $u=u_{+}+\delta u({\bf{r}},t)\exp{(i q.r+\sigma t})$.  The eigenvalue $\sigma$ reads
\begin{equation} \label{Spectrum}
\sigma=\kappa u_{+}(1-2u_{+})-(\delta-\gamma u_{+})q^2-\alpha u_{+}q^4.
\end{equation}
The homogeneous vegetated states exhibit a symmetry-breaking instability at $u=u_{c}$  with an intrinsic wavelength explicitly given by the following simple relation 
\begin{equation} \label{Spectrum}
\lambda=2\pi\sqrt{\frac{2\alpha}{\gamma-\delta/u_{c}}},
\end{equation}
where $u_{c}$ satisfies
\begin{equation} \label{Spectrum}
4\alpha u_{c}^2(2u_{c}-\kappa) =(\gamma u_{c}-\delta)^2.
\end{equation}
This relation determines the threshold state at which the symmetry-breaking instability appears on the
$u_+$ branch of solutions.  We summarize the linear and the weakly nonlinear analysis 
in the bifurcation diagram of Fig. \ref{bifdiam} in  the simplest  monostable case where $\eta<0$. 
A portion of the homogeneously covered vegetation state $u_+$ exhibits a symmetry-breaking 
instability at two thresholds $\eta_c$ and $\eta_c'$ from which several branches of periodic structures emerge.  
The maximum and minimum biomass densities for three kinds of the two-dimensional pattern 
are plotted as a function of the aridity parameter for fixed
values of the cooperativity and the canopy-to-rhizosphere radius ratio. Numerical simulations 
of Eq. (\ref{SH}) supports a well-known scenario: as the aridity increased, the symmetry of 
vegetation patterns transforms from $\pi$-hexagonal into stripes and, finally, into $0$-hexagonal. 
This generic scenario and the associated pattern selection have been predicted in \cite{LEJEUNE99,LEJEUNE04,TLV08}.
Numerical simulations was conducted using the finite differences scheme with Runge-Kutta order-4 algorithm and periodic boundary conditions.

\section{Interaction between localized patches}
The interaction between localized vegetation gaps and patches in arid and semi-arid landscapes has received only scare
attention \cite{TLV08,Berrios}. We show that the interaction between localized vegetation patches is always repulsive \cite{Berrios}. 
The repulsive nature of the interaction between two localized patches do not permit the stabilization of bounded states of vegetation patches. We demonstrate that the repulsive nature of this interaction and the
boundary conditions allow for the coexistence of several vegetation patterns with different wavelengths \cite{Berrios}.
However, in the case of gaps, it has been shown that depending on the distance separating the two spots,
 the interaction  alternates between attractive and repulsive \cite{TLV08}. In next two subsections, 
 we analyze the interaction of two well separated localized gaps in one
and in two-dimensions.

\subsection{Interaction between localized patches in 1D}
\begin{figure}[h]
\begin{center}
\includegraphics[width=0.67\textwidth]{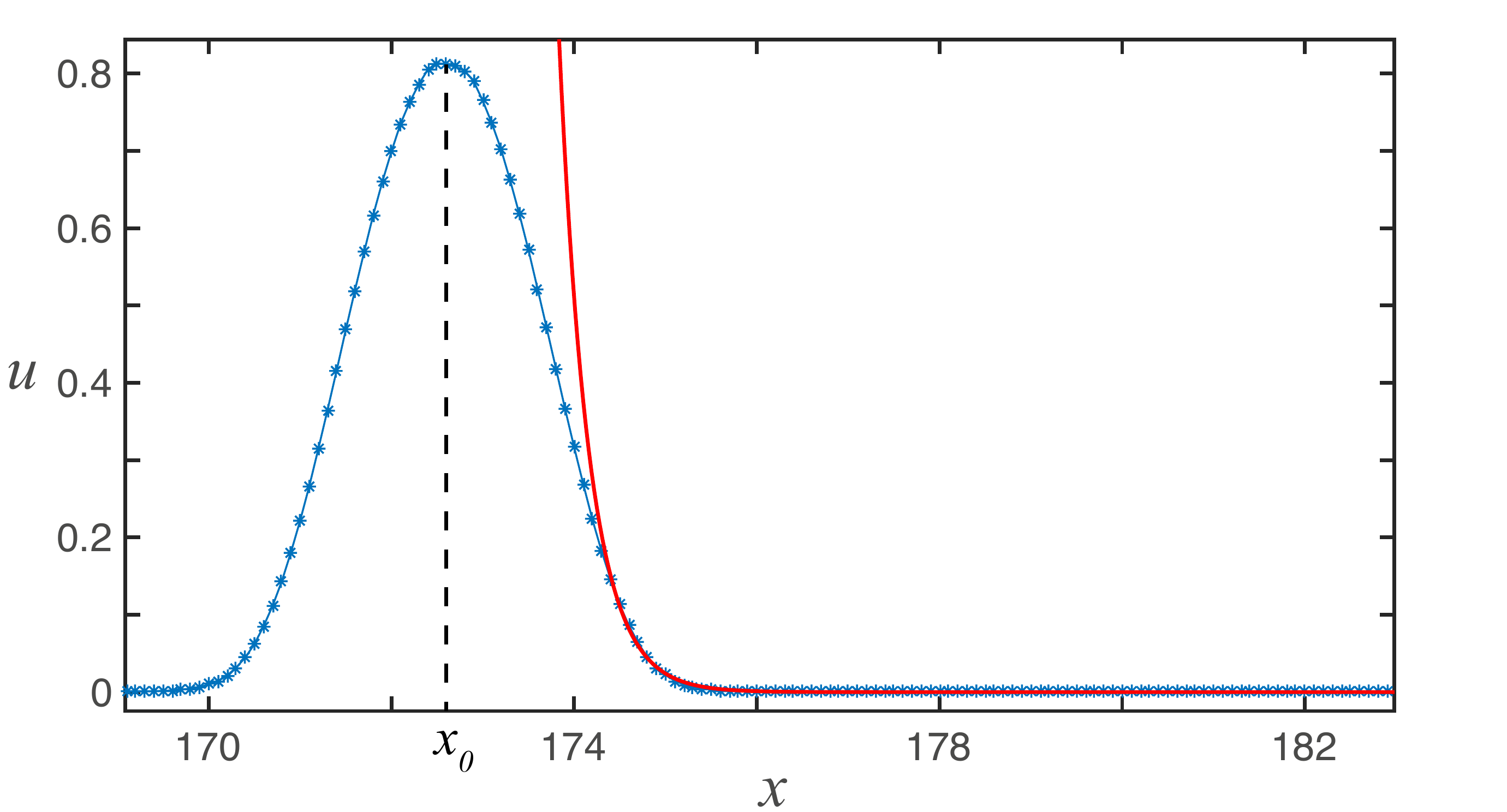}
\end{center}
\caption{One dimensional localized structure profile of the corresponding equation (\ref{SH}). The parameters used are $\eta=0.17, \kappa=0.8, \delta=0.02, \gamma=0.5$ and $\alpha=0.13$. The red line corresponds to the exponential fitting, using Eq. (\ref{uloc-fitting}). For this case the theoretical value of $\gamma$ is 2.91 and from the fitting $3.06$. The coefficient of determination results $R^2=0.9978$.}
\label{LS-asymptotic}
\end{figure}

\begin{figure}[h]
\begin{center}
\includegraphics[width=1\textwidth]{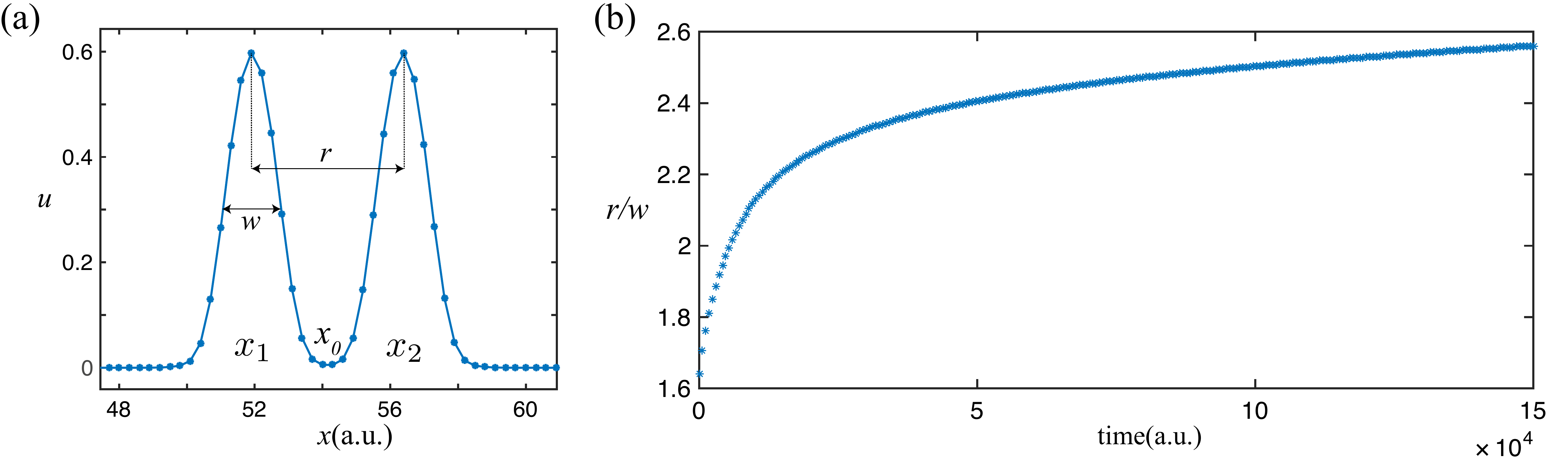}
\end{center}
\caption{(a) Two localized structures separated by a distance $r$. (b) Numerical data of $r$ in function of time, showing the repulsion between LSs. The parameters used are $\eta=0.12, \kappa=0.6, \delta=0.02, \gamma=0.5, \alpha=0.125$ and $dx=0.4$.}
\label{LS-distance}
\end{figure}

To achieve an intuition about the dynamics of localized vegetation patches, 
we will first address the problem of interaction in a one spatial dimension.
A single localized patch $u(x)$ is a solution of  Eq. (\ref{SH}).  The asymptotic behavior of the  tail can be estimated  since localized gap has a small amplitude far from its center, i.e., when $x$ is large, and the asymptotic behavior. We perform a linear analysis around the homogeneous bare sate $u=0$ by linearizing in $u$, we get
$0 = - u \eta + \delta \partial_{xx} u$. The solution of this linear equation is
\begin{equation}
u\left(|x-x_0| \rightarrow  \infty\right){ \mbox{   }} \propto e^{-\beta |x-x_0|}, \qquad    { \mbox{ with  }}  \beta = \sqrt{\eta /\delta},
\label{uloc-fitting}
\end{equation}
where $x_0$ is the center position of localized patch  as depicted 
in  Fig. \ref{LS-asymptotic}. The previous asymptotic behavior  
is confirmed numerically by performing a curve fitting in the tails. 
A single localized patch is a stationary solution of  Eq. (\ref{SH}), which can be interpreted as a nonlinear
front that undergoes a pinning effect between the spatially periodic vegetation and the bare state \cite{Lejeune2002}. The size of an isolated patch is intrinsically determined by the vegetation dynamics and not by some spatial
variation of the environment. It neither grows in spite of available free space, nor decreases in spite of adverse conditions \cite{Lejeune2002}. However, when two localized patches are initially at a certain distance from each other,  they start to move, repelling each other. This repulsion is presented in Fig. \ref{LS-distance} (b), where we have measured numerically the distance of separation $r(t)$ as a  function of time. We have considered $r$ as the distance between the center positions of both interacting patches.
The time evolution of $r$ seems to follow a logarithmic rule, implying that its temporal derivative follows an exponential law in $r$ ($\dot{r} \propto e^{-Ar}$), which makes sense since the asymptotic behavior of the localized structures (LSs) 
tails is exponential. In the next section, we will derive analytically the dynamic equation that $r$ satisfies in a particular limit and will compare with numerical data.

\begin{figure}[h]
\begin{center}
\includegraphics[width=0.7\textwidth]{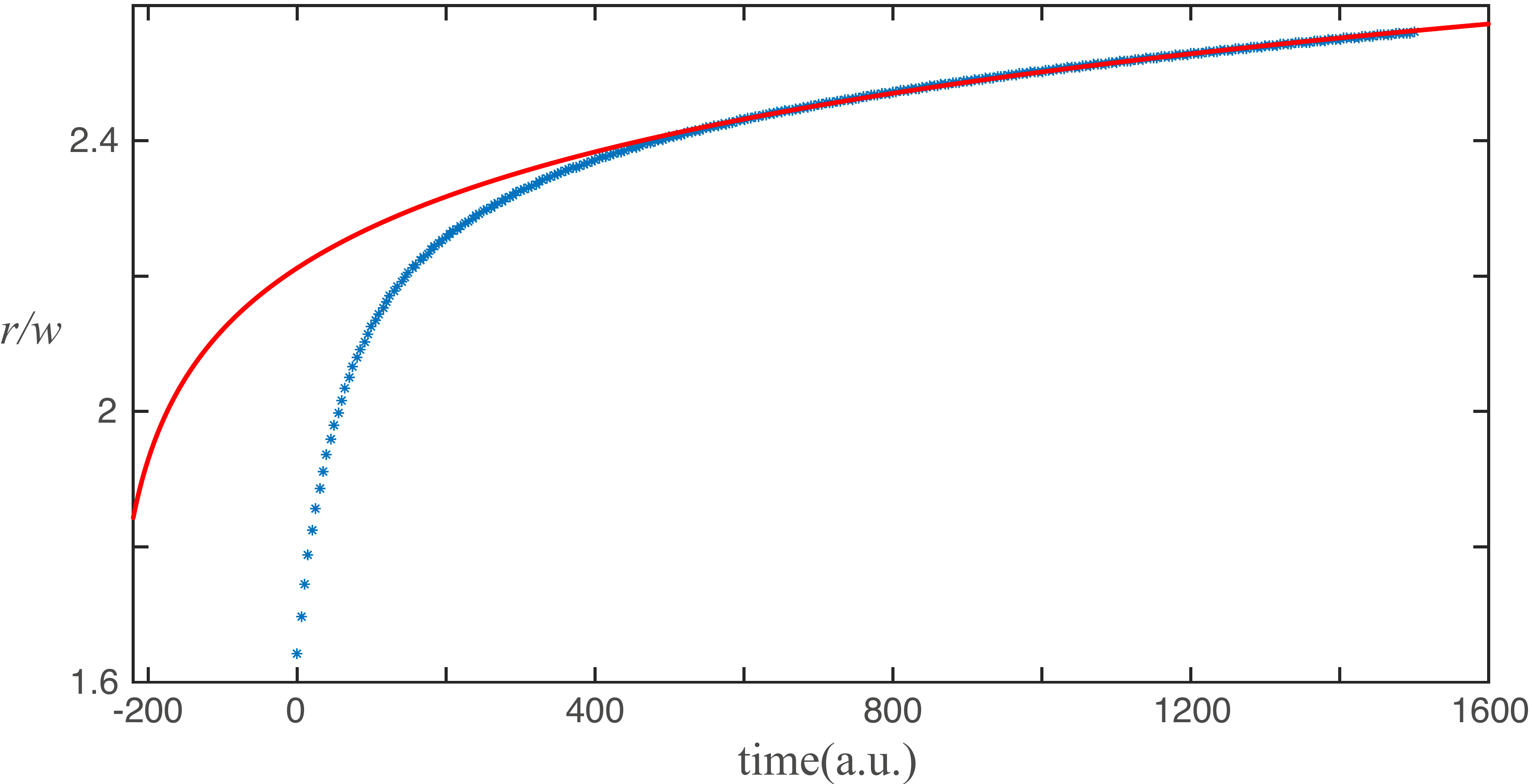}
\end{center}
\caption{Curve fitting of numerical data of $r(t)$. The red curve is the fitting obtained using Eq. (\ref{r-temporal-eq}). The parameters used were $\eta=0.12, \kappa = 0.6, \delta=0.02, \gamma=0.5$, $\alpha=0.125$, $dt=0.01$ and $dx=0.4$. }
\label{distance-fitting}
\end{figure}
In what follows we derive the interaction potential between two localized patches.  For this purpose, we consider a linear superposition of two stationary localized patches $u_{1}$ and $u_{2}$ located at the positions   $x_{1}$ and  $x_{2}$ separate by a distance  $r(t)=x_{2}-x_{1}$
\begin{equation}
u(x,t) = u_{1} (x - x_{1}(t)) + u_{2} (x - x_{2}(t)) + W(x_{1}(t),x_{2}(t),x),
\label{ansatz-1d}
\end{equation}
 where the function $W$ accounts for the corrections due to 
the interaction forces, which will be assumed small. 
The terms proportional to the product of $W$ with $\dot{x}_{-}(t)$ or $\dot{x}_{+}(t)$ also 
will be neglected, since patches move on a slow time scale. In addition, we  
assume that the distance $r$ is large compared to the size of the single localized patch. 
Therefore, introducing this ansatz into the corresponding one dimensional  equation (\ref{SH}), after straightforward calculations (see Appendix A), and by using the solvability condition we get 
\begin{equation}
\partial_t r=A\exp{(-\beta r)}+B\exp{(-2\beta r)}
\label{LSinteraction1D}
\end{equation}
where
\begin{eqnarray}
A&=&\xi \left((\alpha \beta^4+\gamma \beta^2-4\kappa)I_1+3I_2+\gamma I_4+\alpha I_5\right) { \mbox{ and  }}  
B=6 I_3\nonumber \\
\xi&= &<\chi \ \partial_{z_2}u_2-\partial_{z_1}u_1>
\label{Def-AB}
\end{eqnarray}
the integrals $I_j$ with j=1...5 are given in the Appendix 1 [see Eqs. (\ref{eq:I1}-\ref{eq:I5})]

For large $r$, the term $\exp{(-2\beta r)}$ can be neglected. In this limit, Eq. (\ref{LSinteraction1D}) reads
\begin{equation}
\partial_t r=A\exp{(-\beta r)}.
\end{equation}
The solution of this differential equation is
\begin{equation}
 r(t)=\frac{\ln{(A\beta) }}{\beta }+\frac{\ln{(t-t_0)}}{\beta },
\end{equation}
i.e., the distance between two separated localized patches evolves in time according to logarithmic law. This simple formulas are checked with the direct numerical simulation of the governing equation as shown in Fig. \ref{distance-fitting}. The obtained value of tail decay rate associated with localized vegetation patches is $\beta=2.40$, which in a very good agreement with the theoretical value $\beta=(\eta/\delta)^{1/2 }=2.45$. 

\begin{figure}[h]
\begin{center}
\includegraphics[width=1\textwidth]{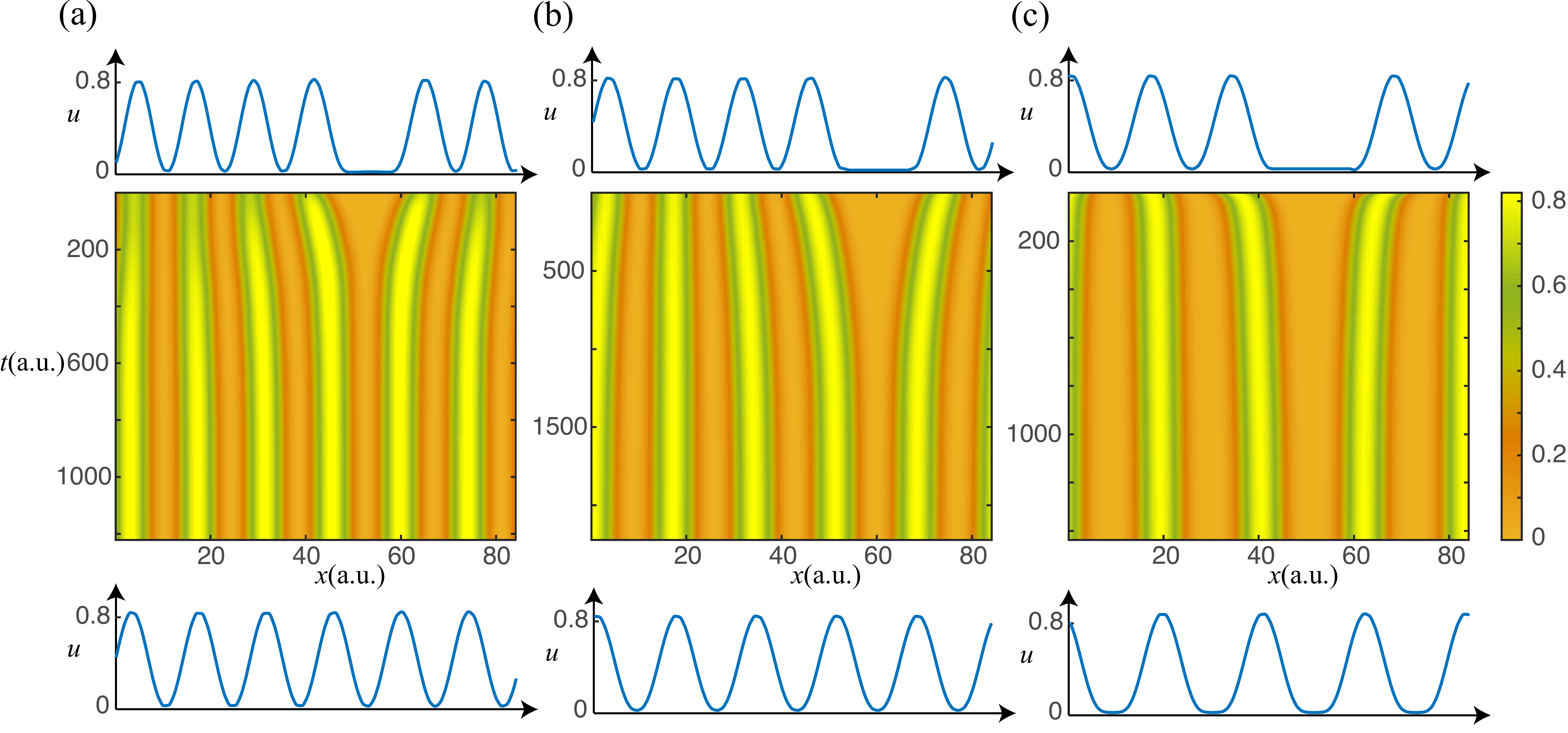}
\end{center}
\caption{Evolution of periodic one-dimensional configurations, after removing one localized structure. The figures (a), (b) and (c) show the evolution of a seven, six, and five  periodic profile evolution, after removing one patch. The upper and lower profiles show the initial and final profile of each case, respectively. In all cases the patches rearrange, reaching a new periodic profile with a larger wavelength. The parameters used were $\eta=0.13, \kappa=0.7, \delta=0.01, \gamma=0.5, \alpha=0.1, dx = 0.26$ and $dt=0.01$.}
\label{veg-evo-periodic-1d}
\end{figure}

In the course of time, the distance between two localized patches increases. Therefore, bounded states of patches are not allowed.  
As a consequence,   the wavelength of periodic train of peaks depends strongly of the system size. 
When a small random initial condition is used, 
in the course of time, the system reaches a periodic structure.
For example, in the left panel of Fig. \ref{veg-evo-periodic-1d}.  If however, we remove one or two peaks, the system will reaches  a stable periodic pattern with bigger wavelength as shown in Fig. \ref{veg-evo-periodic-1d}(b) and (c), respectively.  These figures have been obtained for fixed values of the  parameters, they differ only by the initial conditions. In the next subsection we discuss the interaction in two-dimensional system.

\subsection{Interaction between localized patches in 2D}
\begin{figure}[h]
\begin{center}
\includegraphics[width=0.75\textwidth]{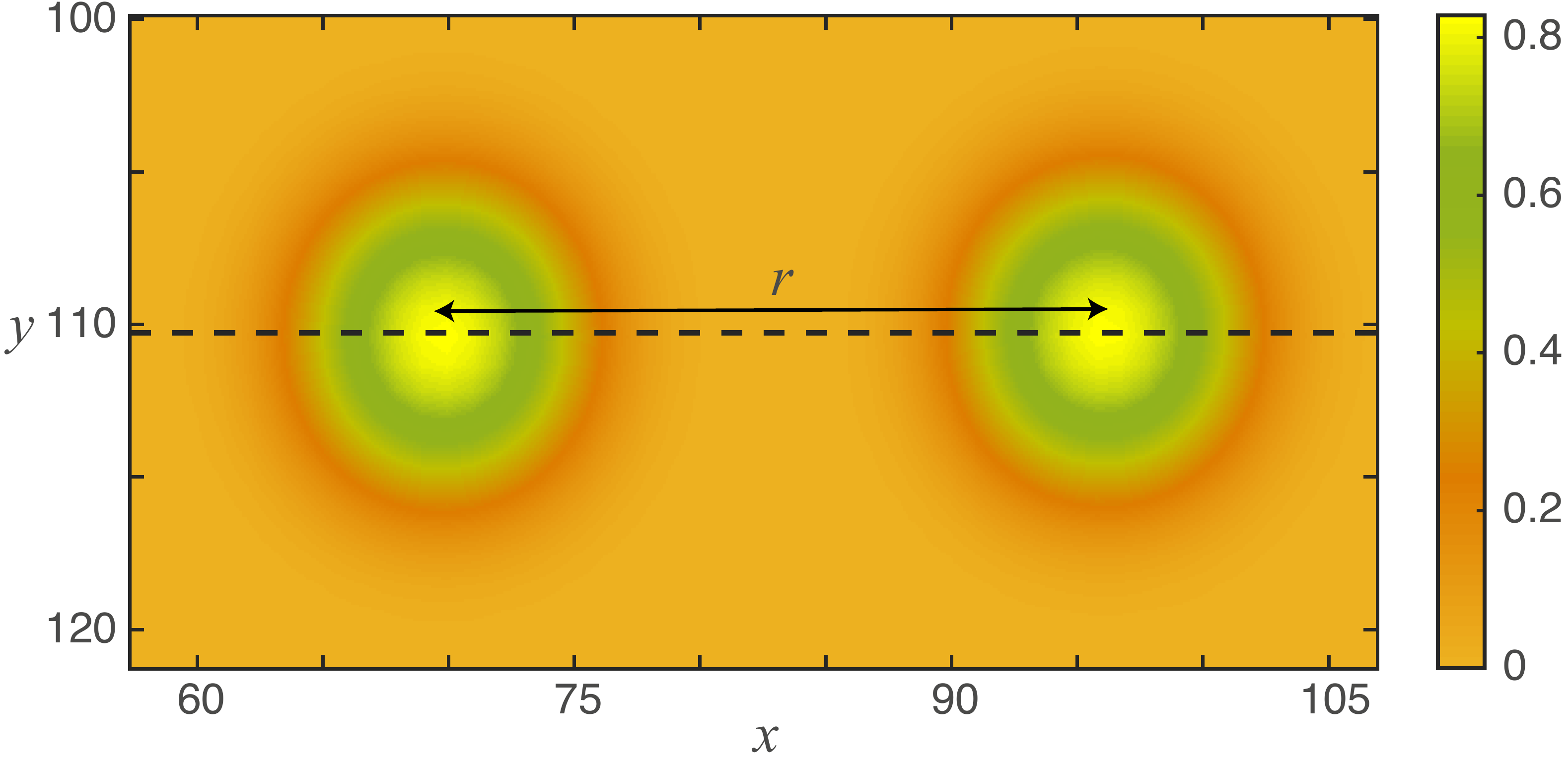}
\end{center}
\caption{Two-dimensional structures located at a distance $r$. 
The dashed line passes through the centers and will be the axis where we restrict our calculations. The
parameters are $\eta=0.12, \kappa=0.6, \delta=0.02, \gamma=0.5, \alpha=0.125$, $dx=dy=0.3$ and $dt=0.001$.}
\label{LS-2D-axis}
\end{figure}

In the previous subsection we have shown that in one dimensional system, the interaction between 
well separated localized patches follows an exponential law. In what follows, 
we focus on 2D interaction problem. 
In this case, the single stationary isolated patch is a solution of the linearized problem around 
the bare state. Since localized solution has a radial symmetry, it is then convenient to express the Laplace operator  in spherical coordinates i.e., $ \nabla^2=1/r\partial_r+\partial^2_r$. The resulting linear problem admits an analytical solution
\begin{equation}
u(r) = A K_0 (\beta r),
\end{equation}
where $A>0$ is a constant, $\beta = \sqrt{\eta/\delta}$, and $K_0$ is the modified Bessel function of second kind, which is a real function for $r>0$. For large values of $r$ we can approximate this function by
\begin{equation}
K_0 (r) \approx \sqrt{\frac{\pi}{2}} \frac{e^{-r}}{\sqrt{r}},
\end{equation}
and then,
\begin{equation}
u_{LS}(r \rightarrow \infty) \propto \frac{e^{-\beta r}}{\sqrt{r}},
\label{asymptotic-2D}
\end{equation}
which describes the asymptotic behavior of the two dimensional localized structures tails. If two or more patches are close one to another they interact througth tair tails. An example of two interacting localized patches is illustrated in 
Fig. \ref{LS-2D-axis}.  
The ansatz we propose for the two interacting patches separated by a distance $r$ is
\begin{equation}
u({\bf x},t) = u_1({\bf x} + r(t)/2) + u_2({\bf x} - r(t)/2) + W(r,{\bf x}),
\end{equation}
 where ${\bf x}$ accounts for the two-dimensional vector, $u_1$ and $u_2$ stand for two localized structures separated 
by a distance $r$, and $W(r,{\bf x})$ is a small correction function. 
Performing an analysis similar to that in the previous section, we obtain
\begin{equation}
\dot{r} = A \frac{e^{- \beta r}}{\sqrt{r}} + B \frac{e^{-2 \beta r}}{r},
\label{dr-2DMAN}
\end{equation}
 where $A$ and $B$ are defined in (\ref{Def-AB}).

\begin{figure}[h]
\begin{center}
\includegraphics[width=1\textwidth]{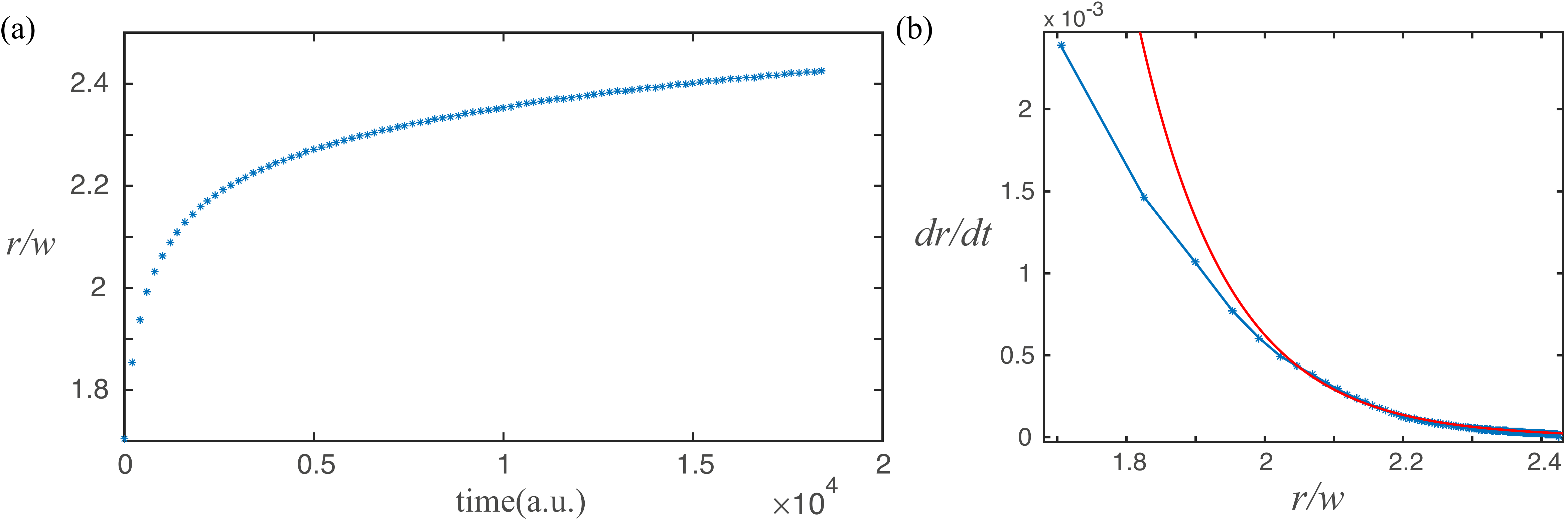}
\label{Fig8}
\end{center}
\caption{(a) Numerical data of the distance of separation $r$ in function of time, in units of the width $w$. (b) Curve fitting of numerical data of $\dot{r}$ in function of $r$, using first term of Eq. (\ref{dr-2D}). The distance $r$ is normalized with the localized patches width. The parameters are the same as in Fig. 7.}
\label{distance-fit-2D}
\end{figure}

Figure~\ref{distance-fit-2D} present a curve fitting of $\dot{r}$ in function of $r$ obtained numerically. The fitting was performed only considering the first term in Eq. (\ref{dr-2DMAN}), assuming that $r$ is large. As is the 1D case, the distance between two interacting patches always increased with time. Therefore, stable 2D bounded patches are unstable since the interaction is always repulsive. 

\section{Interaction between localized gaps}
A single localized gap is a radially symmetric solution $u(x, y) = u(r)$ of Eq. (\ref{SH}). This localized state possesses an oscillatory tail, which can be calculated using the fact that the localized gap amplitude is close to the homogeneous solution $u_+$ far away from its center, i.e., when $r$ is large. Therefore, in order to calculate the asymptotic behavior of the gap solution we substitute into Eq.~(\ref{SH}) an expression $u(\mathbf{r},t)=u_++U(r)$ and linearize this equation around $u_+$:
\begin{equation}
\nabla^2 \left(\begin{array}{c}U\\V\end{array}\right)=M\left(\begin{array}{c}U\\V\end{array}\right)\quad M=\left(\begin{array}{cc}0 & 1\\ -\xi_2& \xi_1\end{array}\right),
\end{equation}
where $V=\nabla^2 U$, $\xi _{1}=\left( \delta  -\gamma u_{+}\right) /\left(\alpha u_{+}\right)$, and $\xi _{2}=(3u_{+}^{2}+\eta -2\kappa u_{+})/(\alpha u_{+})$. The solution of this equation gives the asymptotic behavior of the localized gap in terms of modified Bessel function $K_0$:
\begin{equation}
\left(\begin{array}{c}U\\V\end{array}\right)\approx A\{\vec{v}\exp (i\theta )K_{0}\left[(\omega_1+i \omega_2)r)\right]+c.c.\},  \label{Bessel}
\end{equation}
where
$\omega_1 \pm i\omega_2 =\left[-\xi _{1}/2\pm i\sqrt{\xi _{2}-(\xi _{1}/2)^{2}}\right]^{1/2}$ are complex eigenvalues, while $\vec v=\begin{bmatrix}1 & (\omega_1+i\omega_2)^2\end{bmatrix}^T$ and $\vec v^*=\begin{bmatrix}1 & (\omega_1-i\omega_2)^2\end{bmatrix}^T$ are complex eigenvectors of the matrix $M$. Real constants $A$ and $\theta$ must be calculated numerically.

Two or more localized vegetation gaps will interact through their overlapping tails if they are close enough. The interaction between localized states is a well documented issue in contexts of a physico-chemical \cite{Gorshkov1981,Gorshkov1981,Vladimirov2002,Tlidi_IEEE} rather than biological systems \cite{TLV08}. 

We consider the simplest situation where the two  identical and  radially symmetric interacting gaps are located at the positions  $\mathbf{r}_{1,2}$.  Without the loss of generality we assume that the position of both gaps  are on the  $x$-axis, i.e., their minima are located at the points $(-R/2,0)$ and $(R/2,0)$, where $R=|\mathbf{r}_{2}-\mathbf{r}_{1}|$ is the distance between the gaps. We look for the solution of Eq. (\ref{SH}) in the form of slightly perturbed linear superposition of two gaps:
\begin{equation}
u(\mathbf{r},t)=u_++U_{1}\mathbf{(r)+}U_{2}\mathbf{(r)+}\varepsilon
\delta u(\mathbf{r},t) \quad  \text{where} \quad 
U_{1,2}\mathbf{(r)=}U\left( |\mathbf{r-r}_{1,2}|\right).\label{eq:twogaps}
\end{equation}
The time evolution of the distance between two-identical gaps can be calculated in the regime of week overlap by assuming that the positions  $\mathbf{r}_{1,2}$ evolve on a slow time scale. Substituting Eq.~(\ref{eq:twogaps}) into Eq.~({\ref{SH}) and collecting first order terms in $\varepsilon$ we get an equation with the solvability condition \cite{TLV08}
\begin{equation}
\partial _{t}\mathbf{R}=-\nabla_\mathbf{R}{\cal U}(R),
\end{equation}
 where the potential function
\begin{equation}\label{Potential}
{\cal U}(R)=\alpha u_{+}(-\xi_1 I_{1}+I_{2}).
\end{equation}
and
\begin{eqnarray}
I_{1} &=&-4\int_{-\infty}^{\infty}\left(U
_{x}U\right)_{x=R/2} dy, \label{Int5}\\
 I_{2}&=&-4\int_{-\infty}^{\infty}\left(U V_{x}+U_{x}V\right)_{x=R/2} dy.\label{Int6}
\end{eqnarray}
Here $U_{x}(r)$ is the $x$-component of the eigenfunction of the linear operator $\mathcal{L}^{\dagger}$ adjoint to the operator $\mathcal{L}$, which is  obtained by linearization of  Eq. (\ref{SH}) around the gap solution $U(r)$ and $V(r)=\nabla^2 U(r)$. The asymptotic behaviour of $U_x$ and $V_x=\nabla^2 U_x$ at large $r$ is given by
\begin{equation}
\left(\begin{array}{c}U_x\\V_x\end{array}\right)\approx B\{\frac{x}{r}\vec{v}\exp (i\varphi )K_{1}\left[(\omega_1+i \omega_2)r)\right]+c.c.\} , \label{Bessel1}
\end{equation}
where  real constants $B$ and $\varphi$ must be calculated numerically. A detailed derivation of Eq. (\ref{Potential}) as well as the evaluation of the integrals $I_{1,2}$ can be found in the  book  chapter \cite{TLV08}. 
After substituting asymptotic relations (\ref{Bessel}) and (\ref{Bessel1}) into Eqs. (\ref{Int5}) and (\ref{Int6}), and performing integration we finally get: 
\begin{equation}
{\cal U}(R)=-8\pi A B \alpha u_+\Im\{\frac{\omega_1\omega_2e^{i(\theta+\varphi)}}{(\omega_1+i \omega_2)^2} K_0\left[(\omega_1+i\omega_2)R\right]\}.
\end{equation}
The minima (maxima) of the interaction potential correspond to stable (unstable) localized gaps bound states, where two bare spots are bounded together by  the interaction force. The interaction potential associated with two-gaps is plotted in Fig.~\ref{FIG9}.
\begin{figure}
\label{FIG9}
\centering\includegraphics[width=9cm]{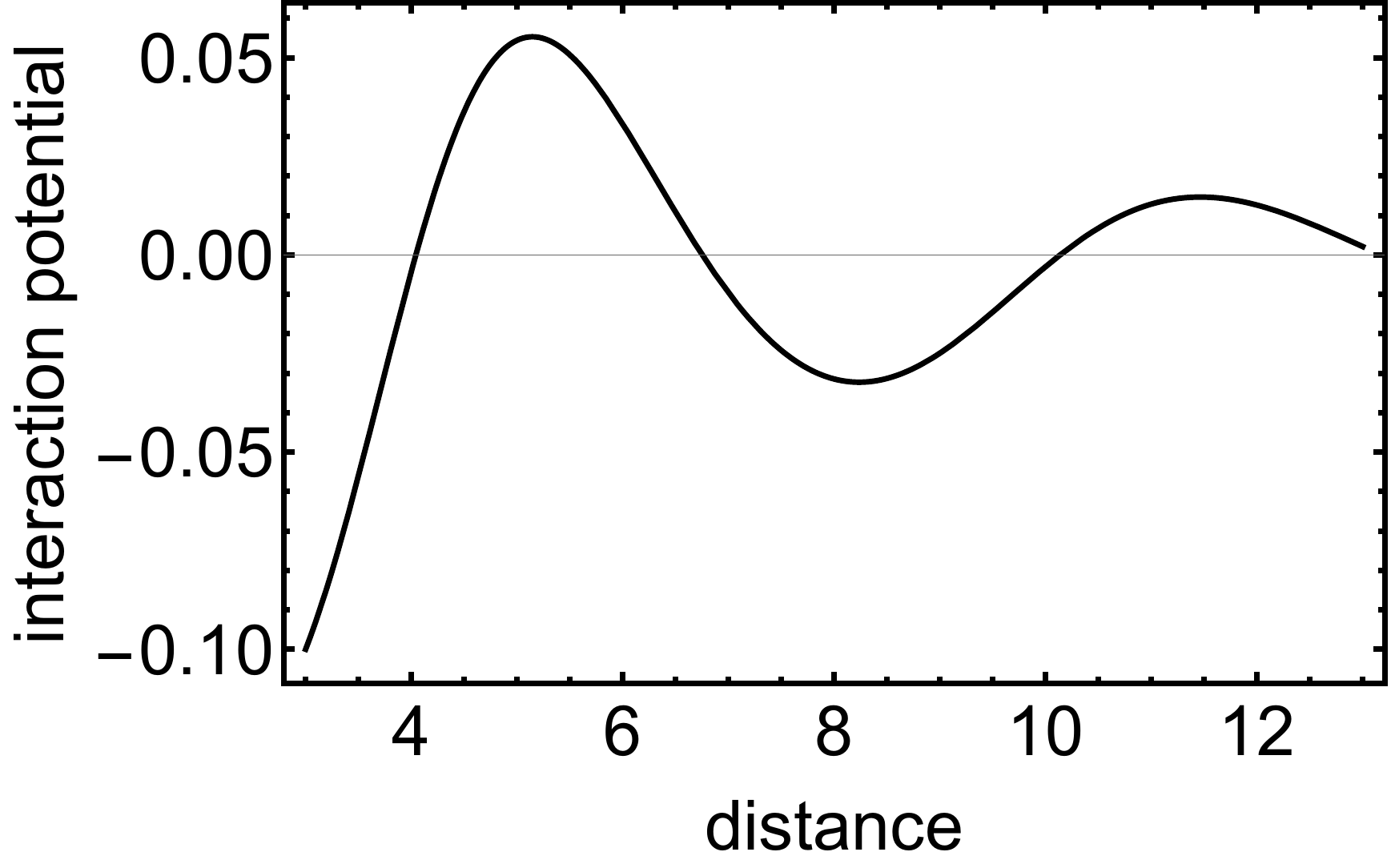}
\caption{Interaction potential as a function of between two localized gaps. Parameters are $\kappa=0.5$, $\delta=0.1$,$\gamma=2\alpha=0.5$, and $\eta=-0.025$}
\end{figure}

\section{Conclusions}
Starting from the the propagation-inhibition model, we have generalized the derivation of a real partial differential equation for weak Kernels such as Gaussian or exponential. This derivation is valid in the double limit: close to the critical point associated with bistability and close to the long wavelength pattern forming regime. The Eq. (\ref{SH}) differs from the usual Swift-Hohenberg equation in the fact that the terms $u\nabla^2u$ and $u\nabla^4u$ render this model nonvarialional, i.e., there is no potential or Lyapunov fonctional to minimize near the double limit we consider.  In last part, we have focused our analysis of the interaction between two well separated gaps solutions of Eq. (\ref{SH}). We have constructed the interaction potential. We have shown that in this case, the interaction alternates between attractive and repulsive depending on the distance
separating the gaps. The existence of discrete stable equilibrium position allows for the stabilization of bounded states and clusters of gaps. However, in the case of localized patches, we have shows the interaction is always repulsive. Therefore, on the long time scale bounded localized patches are excluded. \\
{\bf{Acknowledgments}}\\
M.T. is a Research Director with the Fonds de la Recherche Scientifique F.R.S.-FNRS, Belgium. 
MGC and DPR thank financial support of  the Millennium Institute for Research in Optics (Miro).

\appendix
\section{Derivation of the interaction potential in 1D}

We derive the interaction law of two localized structures separated by a distance $r(t) = x_2(t) - x_1(t)$, with the central position 
or centroid  $x_0 = (x_2(t) - x_1(t))/2$ [see Fig.~\ref{LS-distance} (a)]. 
Let us  first introduce the ansatz presented in Eq.~(\ref{ansatz-1d}) into Eq.~(\ref{SH}), 
i.e. the dynamical equation of the system, and derive a linear system in $W$, the correction term. Recall that $x_1(t)$ and $x_2(t)$ 
account for positions of structures $u_1$ and $u_2$, respectively. Notice that they have been promoted as functions of time. Our aim in this section is to derive dynamical equations for $r(t)$ and $x_0(t)$ under certain assumptions.

As mentioned in the main text, it is assumed that the term $W$ is small, i.e. nonlinear terms are neglected. Moreover, the localized structure is assumed to travel slow enough so that the terms proportional to the product of $W$ with either $\dot{x}_1$ or $\dot{x}_2$ are not considered. Therefore, under the previous assumptions, the linear system associated to $W$ becomes
\begin{equation}
 \mathcal{L} W = b,
 \label{linear-system-L}
\end{equation}
with the linear operator $\mathcal{L}$ defined as
\begin{align}
  &\begin{aligned}
\mathcal{L }= & - \eta + 2\kappa(u_1 + u_2) - 3(u_1 + u_2)^2 + \delta \partial_{xx}\\
&  - \gamma \left[(u_1 + u_2) \partial_{xx} +  \partial_{xx} (u_1 + u_2) \right] \\
&  - \alpha \left[(u_1 + u_2) \partial_{xxxx} +  \partial_{xxxx} (u_1 + u_2) \right],
  \end{aligned}
\label{expansion-magnetic-2}
\end{align}
and
\begin{align}
  &\begin{aligned}
b = & \dfrac{\dot{r}}{2} \left(\partial_{z_1} u_1 - \partial_{z_2} u_2 \right) - \dot{x}_0 \left(\partial_{z_1} u_1 + \partial_{z_2} u_2 \right) \\
&  -2 \kappa  u_1 u_2 + 3 u_1 u_2 \left( u_1 + u_2 \right)\\
& + \gamma \left(u_1 \partial_{xx} u_2 + u_2 \partial_{xx} u_1 \right)\\
& + \alpha \left(u_1 \partial_{xxxx} u_2 + u_2 \partial_{xxxx} u_1 \right).\\
  \end{aligned}
  \label{b-1D}
\end{align}
In our calculations, we have used the fact that $u_1$ and $u_2$ are solutions of the dynamical equation of the system.

From here, the strategy to characterize the dynamics of $r(t)$ and $x_0(t)$ will be by applying the Fredholm solvability condition \cite{Fredholm} to this system, for the following inner product
\begin{equation}
\braket{f | g} = \int_{- \infty}^{+ \infty} f(x) g(x) \: \mathrm{d} x,
\end{equation}
with $f(x)$ and $g(x)$ as real-valued functions. The derivation of the associated adjoint operator $\mathcal{L}^{\dagger}$ is necessary for applying this condition.

\subsection{Adjoint of $\mathcal{L}$}

To obtain $\mathcal{L}^{\dagger}$, we take a look at the property $\braket{\mathcal{L}^{\dagger} f | g}=\braket{f | \mathcal{L}g}$ for adjoint operators. From here, it is straightforward to see that the terms of $\mathcal{L}$ remain the same as in $\mathcal{L}^\dagger$, except
\begin{equation}
 \gamma (u_1 + u_2) \partial_{xx} \quad \text{and} \quad \alpha (u_1 + u_2) \partial_{xxxx}.
 \label{terms-adjoint}
\end{equation}
The adjoint of these terms can be obtained by integrating by parts. We illustrate this procedure for the first of them. 

When introducing $ \gamma (u_1 + u_2) \partial_{xx} $ into the inner product
\begin{equation}
\braket{f | \gamma (u_1 + u_2) \partial_{xx}g} = \int_{- \infty}^{+ \infty} f(x) \: \gamma \left[u_1(x) + u_2(x) \right] \partial_{xx}g(x) \: \mathrm{d} x,
\end{equation}
and integrating by parts once, we get
\begin{equation}
f(x) \: \gamma \left[u_1(x) + u_2(x) \right] \partial_{xx}g(x)  \Big|_{-\infty}^{\infty}
- \int_{- \infty}^{+ \infty}  \partial_{x}[f(x) \: \gamma (u_1(x) + u_2(x))] \partial_{x}g(x) \: \mathrm{d} x.
\end{equation}
The time dependence of $u_1$ and $u_2$ has been suppressed for simplicity of notation. 
The first of these terms vanishes, as $(u_1 + u_2)$ tends to zero as $x \rightarrow \pm \infty$. 
Integrating by parts once more, and proceeding in a similar way, we get 
\begin{equation}
\braket{f | \gamma (u_1 + u_2) \partial_{xx}g} = \braket{ \partial_{xx}[f \: \gamma (u_1 + u_2)] | g},
\end{equation}
so that, the corresponding element of $\gamma (u_1 + u_2) \partial_{xx}$ in $\mathcal{L}^\dagger$ 
takes the form $\partial_{xx}[\gamma (u_1 + u_2) \cdot]$. 
Notice that operator receives arguments inside the partial derivative $\partial_{xx}$. 
This is represented with a dot inside the square brackets.

Proceeding in the same way with the second term in Eq. (\ref{terms-adjoint}), 
the adjoint operator becomes
\begin{align}
  &\begin{aligned}
\mathcal{L}^{\dagger} = & - \eta + 2\kappa(u_1 + u_2) - 3(u_1 + u_2)^2 + \delta \partial_{xx}\\
&  - \gamma \left\{(u_1 + u_2) \partial_{xx} +  \partial_{xx} \left[(u_1 + u_2) \cdot \right] \right\}\\
& - \alpha \left\{(u_1 + u_2) \partial_{xxxx} +  \partial_{xxxx} \left[(u_1 + u_2) \cdot \right]\right\}.\\
  \end{aligned}
\end{align}

\subsection{Kernel of $\mathcal{L}^{\dagger}$}

To apply the Fredholm solvability condition, it is necessary to calculate 
the Kernel components of $\mathcal{L}^{\dagger}$, that is, the elements that fulfill 
$\bra{f} \mathcal{L}^{\dagger} = 0$. Due to the complexity of $\mathcal{L}^{\dagger}$, 
they are obtained numerically by discretizing its derivatives, using central differencing 
with the 4 nearest neighbors. Thus, when applying $\mathcal{L}^{\dagger}$ 
to a real-values function $f$, the discrete form takes the form
\begin{align}
  &\begin{aligned}
\bra{f_j} \mathcal{L}^{\dagger}  = & \left(\frac{7 c_4}{240 \text{dx}^4}-\frac{7 c_3}{240 \text{dx}^3}-\frac{c_2}{560 \text{dx}^2}+\frac{c_1}{280 \text{dx}}\right) f_{j-4} \\
& +\left(-\frac{2 c_4}{5 \text{dx}^4}+\frac{3 c_3}{10 \text{dx}^3}+\frac{8 c_2}{315 \text{dx}^2}-\frac{4 c_1}{105 \text{dx}}\right) f_{j-3} \\
& +\left(\frac{169 c_4}{60 \text{dx}^4}-\frac{169 c_3}{120 \text{dx}^3}-\frac{c_2}{5 \text{dx}^2}+\frac{c_1}{5 \text{dx}}\right) f_{j-2} \\
& +\left(-\frac{122 c_4}{15 \text{dx}^4}+\frac{61 c_3}{30 \text{dx}^3}+\frac{8 c_2}{5 \text{dx}^2}-\frac{4 c_1}{5 \text{dx}}\right) f_{j-1} \\ 
& + \left(\frac{91 c_4}{8 \text{dx}^4}-\frac{205 c_2}{72 \text{dx}^2}+c_0\right) f_{j} \\
& +\left(-\frac{122 c_4}{15 \text{dx}^4}-\frac{61 c_3}{30 \text{dx}^3}+\frac{8 c_2}{5 \text{dx}^2}+\frac{4 c_1}{5 \text{dx}}\right) f_{j+1} \\
& +\left(\frac{169 c_4}{60 \text{dx}^4}+\frac{169 c_3}{120 \text{dx}^3}-\frac{c_2}{5 \text{dx}^2}-\frac{c_1}{5 \text{dx}}\right) f_{j+2} \\
& +\left(-\frac{2 c_4}{5 \text{dx}^4}-\frac{3 c_3}{10 \text{dx}^3}+\frac{8 c_2}{315 \text{dx}^2}+\frac{4 c_1}{105 \text{dx}}\right) f_{j+3} \\
& +\left(\frac{7 c_4}{240 \text{dx}^4}+\frac{7 c_3}{240 \text{dx}^3}-\frac{c_2}{560 \text{dx}^2}-\frac{c_1}{280 \text{dx}}\right) f_{j+4},
  \end{aligned}
\end{align}
with
\begin{align}
  &\begin{aligned}
c_0 = & -\eta + 2 \kappa U - 3U^2 - 2 \gamma \partial_{xx} U - 2 \alpha \partial_{xxxx} U,\\
c_1 = & - 2 \gamma \partial_x U - 4 \alpha \partial_{xxx} U,\\
c_2 = & \: \delta - \gamma U - 6 \alpha \partial_{xx} U,\\
c_3 = & - 4 \alpha \partial_x U,\\
c_4 = & - \alpha U \partial_{xxxx}, \\
  \end{aligned}
\end{align}
where $U = u_1 + u_2$, and $f_j \equiv f(x = \mathrm{d}x j)$, with $dx$ the discretization used. 

Finding the Kernel of $\mathcal{L}^{\dagger}$ is equivalent to determine the Kernel (or nullspace) of a matrix $M$ that satisfies
\begin{equation}
M \vec{f} = \vec{0} \quad \text{with} \quad \vec{f} = \left( \begin{array}{c}
f_1 \\
\vdots \\
f_{j-1} \\
f_{j} \\
f_{j+1} \\
\vdots \\
f_{N}\\
\end{array}
\right),
\label{system-M-f}
\end{equation}
with $N$ the number of points considered in the discretization. Let us first take a look at the eigenvalue 
spectrum of $M$. Using $\mathrm{d}x =0.1$, $N=30$, and setting $u_{LS}^-$ and $u_{LS}^+$ 
at a distance $r=150$, we obtain the eigenvalue spectrum shown in Fig. \ref{spectrum-eigenvalues} 
(a) and (b). As the graph shows, we confirm the stability of the localized structures as every 
eigenvalue has a negative real part. Moreover, the lowest eigenvalues ($-0.022$ and $-0.024$) 
decrease as the distance $r$ and the number of points $N$ increase, implying that they 
correspond to the null eigenvalues we are looking for. 
They are not exactly equal to zero due to the numerical approximation.

Finally, the linear combination of the eigenvectors associated with the calculated null eigenvalues 
gives us the elements of the Kernel of $\mathcal{L}^{\dagger}$. 
These will be labeled as $\bra{\tau}$ and $\bra{\chi}$. 
Their respective profiles are shown in Fig. \ref{spectrum-eigenvalues} (c) and (d).

\begin{figure}[h]
\begin{center}
\includegraphics[width=0.85\textwidth]{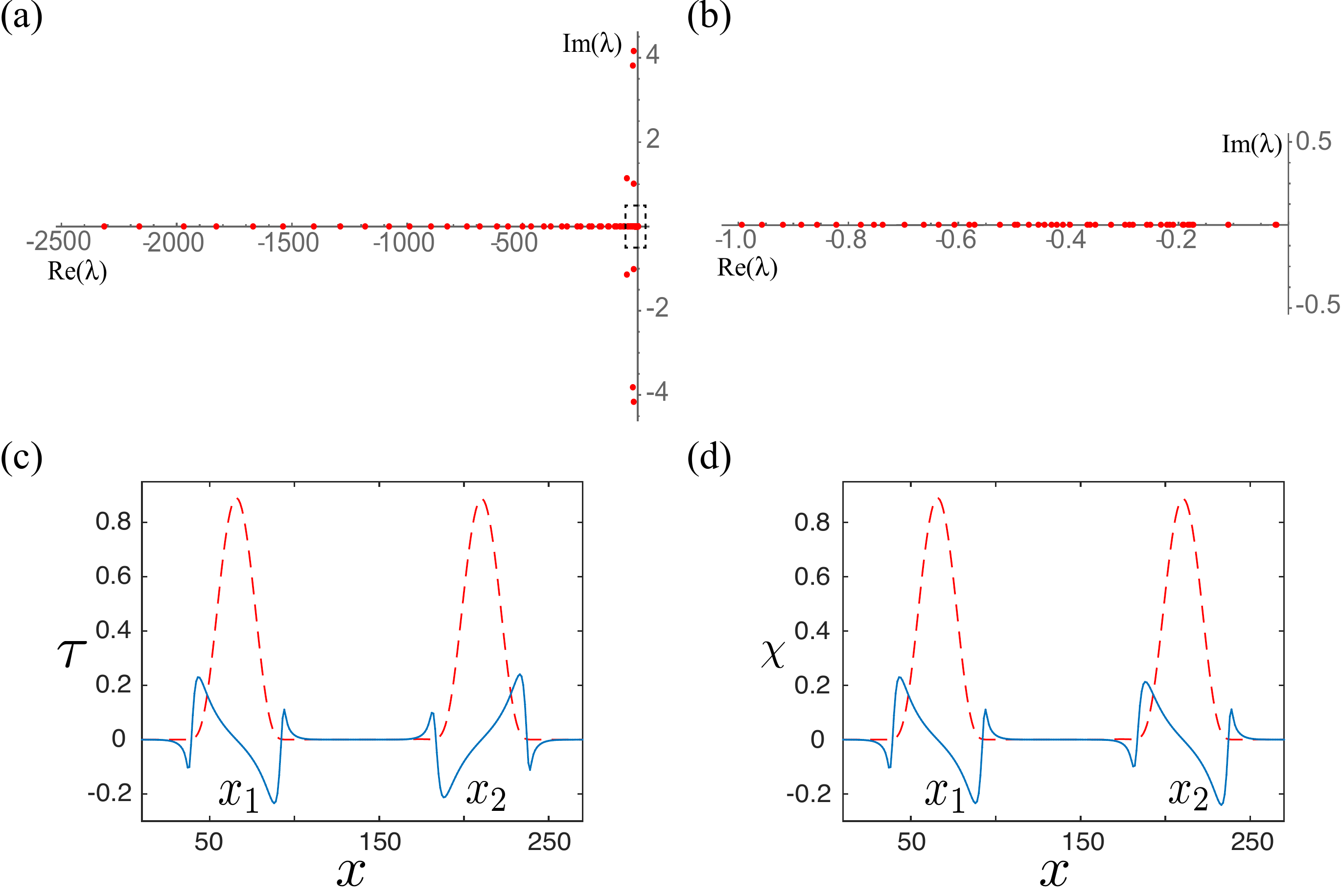}
\end{center}
\caption{(a) Eigenvalue spectrum of matrix $M$ using $\mathrm{d}x =0.1$ and $N=30$. (b) Zoom of the dashed region marked in (a). Figures (c) and (d) show the null eigenvector $\tau$ and $\chi$ (solid line), respectively, and the associated localized structures profile (dashed line). Parameters used are the same as in Figure \ref{LS-distance}.}
\label{spectrum-eigenvalues}
\end{figure}

\subsection{Interaction dynamical equations}

Having the Kernel of $\mathcal{L}^{\dagger}$, the dynamical equations of $r$ and $x_0$ are determined by applying the Fredholm solvability condition. This states that the linear system (\ref{linear-system-L}) admits solution if and only if
\begin{equation}
\braket{\tau | b}=0
\label{fredholm-tau-veg}
\end{equation}
and
\begin{equation}
\braket{\chi | b}=0.
\label{fredholm-chi-veg}
\end{equation}

In what follows, we expand both products and obtain dynamical equations for both $x_0(t)$ and $r(t)$.

\subsubsection{Equation for the central position $x_0(t)$}

In the first of these products, Eq. (\ref{fredholm-tau-veg}), given the form of $b$ (see Eq. (\ref{b-1D})), and since $\tau$ is odd around $x_0$, the only term that remains is
\begin{equation}
\braket{\tau | \dot{x}_0 \left(\partial_{z_1} u_1 + \partial_{z_2} u_2 \right)} = 0,
\end{equation}
implying that
\begin{equation}
\dot{x}_0 = 0,
\end{equation}
i.e., the central position of the LSs does not move. This result is observed in the numerical simulations, which is expected due to the symmetries of the dynamical equation of the system. 

\subsubsection{Equation for the distance $r(t)$}

On the other hand, in the second product, Eq. (\ref{fredholm-chi-veg}), due to the symmetries of the terms involved, the ones that remain are
\begin{align}
  &\begin{aligned}
& \braket{\chi | (\dot{r}/2) \left(\partial_{z_1} u_1 - \partial_{z_2} u_2 \right)} +  \\
&  \bra{\chi}-2 \kappa  u_1 u_2 + 3 u_1 u_2 \left( u_1 + u_2 \right) \\
& + \gamma \left(u_1 \partial_{xx} u_2 + u_2 \partial_{xx} u_1 \right)\\
& + \alpha \left(u_1 \partial_{xxxx} u_2 + u_2 \partial_{xxxx} u_1 \right) \rangle = 0.\\
  \end{aligned}
  \label{chi-products-veg}
\end{align}
The integrals involved in this equation can be approximated analytically. For this, it is convenient to write
\begin{equation}
\chi(x) = \chi_- (x - x_1) + \chi_+ (x - x_2),
\label{split-chi}
\end{equation}
that is, to divide $\chi$ into two parts, one localized around $x_1$ and the other around $x_2$. To illustrate how to approximate the integrals analytically, let us take a look at the second product of (\ref{chi-products-veg}), i.e.
\begin{align}
  &\begin{aligned}
& \braket{\chi | 2 \kappa u_1 u_2}  =   2 \kappa \int_{-\infty}^{\infty} \left[ \chi^{-}(x - x_1) + \chi^{+} (x - x_2) \right] u_1(x - x_1) u_2(x - x_2) \mathrm{d} x \\
& = 2 \kappa \left[ \int_{-\infty}^{\infty} \chi^{-}(z_1) u_1(z_1) u_2(z_1 - r) \mathrm{d} z_1 +  \int_{-\infty}^{\infty} \chi^{+}(z_2) u_1(z_2 + r) u_2(z_2) \mathrm{d} z_2 \right].\\
  \end{aligned}
  \label{inner-product-chi}
\end{align}
In the second equality, we have changed our variables to $z_1 = x - x_1$ and $z_2 = x - x_2$ in the first and second integrals, respectively. We have also used $r = x_2 - x_1$. These integrals are exponentially close to zero in the whole region of integration, except when they are evaluated near zero. Thus, a good approximation of them is given by setting the integral limits from $-r/2$ to $r/2$. Moreover, since $r$ is large, the terms $u_2(z_1 - r)$ and $u_1(z_2 + r)$ are exponentially small in the region of integration. An approximation of these functions is given then by the asymptotic behavior of $u_1$ and $u_2$ (see Eq. (\ref{uloc-fitting})), so that
\begin{align}
  &\begin{aligned}
& \braket{\chi | 2 \kappa u_1 u_2} \approx \\
& 2 \kappa \left[ \int_{-r/2}^{r/2} \chi^{-}(z_1) u_1(z_1) e^{-\beta |z_1 - r|} \mathrm{d} z_1 +  \int_{-r/2}^{r/2} \chi^{+}(z_2)  e^{-\beta |z_2 + r|}  u_2(z_2) \mathrm{d} z_2 \right].\\
  \end{aligned}
\end{align}
 where $|z_1 - r| = r - z_1$ and $ |z_2 + r| = z_2 + r$ in this region of integration. We write
\begin{equation}
\braket{\chi | - 2 \kappa u_1 u_2}   \approx - 2  \kappa e^{-\beta r} I_1,
\end{equation}
where
\begin{equation}
I_1 = -\left[ \int_{-r/2}^{r/2} \chi^{-}(z_1) u_1(z_1) e^{\beta z_1} \mathrm{d} z_1 +  \int_{-r/2}^{r/2} \chi^{+}(z_2) e^{-\beta z_2} u_2(z_2) \mathrm{d} z_2 \right].\label{eq:I1}
\end{equation}

Proceeding in the same way with the other integrals in Eq (\ref{chi-products-veg}), we finally obtain the dynamical equation for $r$
\begin{equation}
\dot{r} = A e^{- \beta r} + B e^{-2 \beta r},
\label{eq-dr-full}
\end{equation}
where
\begin{equation}
A =  \dfrac{2 \left[ (\alpha \beta^4 + \gamma \beta^2 - 2\kappa) I_1  + 3 I_2 + \gamma I_4 + \alpha I_5 \right]}{\braket{\chi | \partial_{z_1} u_1 - \partial_{z_2} u_2 }}
\end{equation}
and
\begin{equation}
B = \dfrac{6 I_3}{\braket{\chi | \partial_{z_1} u_1 - \partial_{z_2} u_2}},
\end{equation}
with
\begin{equation}
I_2 = -\left[ \int_{-r/2}^{r/2} \chi^{-}(z_1) (u_1(z_1))^2 e^{\beta z_1} \mathrm{d} z_1 +  \int_{-r/2}^{r/2} \chi^{+}(z_2) e^{-\beta z_2} (u_2(z_2))^2 \mathrm{d} z_2 \right],
\end{equation}
\begin{equation}
I_3 = -\left[ \int_{-r/2}^{r/2} \chi^{-}(z_1) u_1(z_1) e^{2 \beta z_1} \mathrm{d} z_1 +  \int_{-r/2}^{r/2} \chi^{+}(z_2) e^{-2\beta z_2} u_2(z_2) \mathrm{d} z_2 \right],
\end{equation}
\begin{equation}
I_4 = -\left[ \int_{-r/2}^{r/2} \chi^{-}(z_1) \partial_{z_1 z_1}u_1(z_1) e^{\beta z_1} \mathrm{d} z_1 +  \int_{-r/2}^{r/2} \chi^{+}(z_2) e^{-\beta z_2}  \partial_{z_2 z_2}u_2(z_2) \mathrm{d} z_2 \right]
\end{equation}
and
\begin{equation}
I_5 = -\left[ \int_{-r/2}^{r/2} \chi^{-}(z_1)  \partial_{z_1}^{(4)} u_1(z_1) e^{\beta z_1} \mathrm{d} z_1 +  \int_{-r/2}^{r/2} \chi^{+}(z_2) e^{-\beta z_2}  \partial_{z_2}^{(4)} u_2(z_2) \mathrm{d} z_2 \right].\label{eq:I5}
\end{equation}

All these integrals can be calculated numerically. We have checked that factors $A$ and $B$ are positive for the range of parameters used in the simulations. In addition, since $r$ is large, we neglect the $e^{- 2 \beta r}$ term in Eq. (\ref{eq-dr-full}) and write as
\begin{equation}
\dot{r} = A e^{-\beta r},
\end{equation}
from where we can derive the temporal dependence of $r$
\begin{equation}
r(t) = \dfrac{1}{\beta} \ln (t-t_0) + \dfrac{1}{\beta} \ln (A \beta),
\label{r-temporal-eq}
\end{equation}
which agrees with the numerical data, as shown in the curve fitting in Fig. \ref{distance-fitting}, for large $r$. In fact, the fitting is better when the distance of separation is more than $2.3$ times the localized structures width $w$, for the parameters used in Figure \ref{LS-distance}. The $\beta$ factor obtained from the fitting is in good agreement with the value used in the simulations.

\section{Derivation of the interaction potential in 2D}

In this section, we derive the interaction law for $r(t)$ in two dimensions. The procedure is essentially the same as in one dimension, as we focus the dynamics on the $x$ axis. The main difference relies on the asymptotic behavior of the localized structures. In fact, the linear system obtained after replacing the ansatz will be 
\begin{equation}
\mathcal{L} W = b,
\end{equation}
with the same operator $\mathcal{L}$ as in one dimension, and
\begin{align}
  &\begin{aligned}
b = & \dfrac{\dot{r}}{2} \left(\partial_{z_1} u_1 - \partial_{z_2} u_2 \right) \\
&  -2 \kappa  u_1 u_2 + 3 u_1 u_2 \left( u_1 + u_2 \right)\\
& + \gamma \left(u_1 \partial_{xx} u_2 + u_2 \partial_{xx} u_1 \right)\\
& + \alpha \left(u_1 \partial_{xxxx} u_2 + u_2 \partial_{xxxx} u_1 \right),\\
  \end{aligned}
\end{align}
with $z_\pm = x \mp r/2$. We choose the same inner product as in one dimension, and in consequence, we obtain the same Kernel of $\mathcal{L}^\dagger$. These are labeled again as $\bra{\tau}$ and $\bra{\chi}$.

By applying the Fredholm solvability condition, $\braket{\tau | b}=0$, we obtain that $\dot{x}_0 = 0$, i.e. there is no dynamics in the center position, as in one dimension. For the second product, $\braket{\chi | b}=0$, the analytical approximations needed are slightly different than in one dimension. To illustrate them, let us consider the inner product between $\chi$ and $2 \kappa u_1 u_2$ (as in Eq. (\ref{inner-product-chi}))
\begin{align}
  &\begin{aligned}
& \braket{\chi | 2 \kappa u_1 u_2}  \\
& = 2 \kappa \left[ \int_{-\infty}^{\infty} \chi^{-}(z_1) u_1(z_1) u_2(z_1 - r) \mathrm{d} z_1 +  \int_{-\infty}^{\infty} \chi^{+}(z_2) u_1(z_2 + r) u_2(z_2) \mathrm{d} z_2 \right].\\
  \end{aligned}
\end{align}
Again, we have split $\chi$ as in Eq. (\ref{split-chi}). We also restrict the integral limits only from $-r/2$ to $r/2$ and replace $u_2(z_1 - r)$ and $u_1(z_2 + r)$ by their asymptotic behavior . Since now we are focusing on the dynamics on the $x$ axis, the asymptotic behavior takes the form (see Eq. (\ref{asymptotic-2D}))
\begin{equation}
u_{LS}(|x-x_0| \rightarrow \infty) \propto \dfrac{e^{-\beta |x-x_0|}}{\sqrt{|x-x_0|}},
\end{equation}
where $x_0$ is the central position. Thus, 
\begin{align}
  &\begin{aligned}
& \braket{\chi | 2 \kappa u_1 u_2} \approx \\
& 2 \kappa \left[ \int_{-r/2}^{r/2} \chi^{-}(z_1) u_1(z_1) \dfrac{e^{-\beta |z_1 - r|}}{\sqrt{|z_1 - r|}} \mathrm{d} z_1 +  \int_{-r/2}^{r/2} \chi^{+}(z_2)  \dfrac{e^{-\beta |z_2 + r|}}{\sqrt{|z_2 + r|}}  u_2(z_2) \mathrm{d} z_2 \right].\\
  \end{aligned}
\end{align}
Notice that $|z_1 - r| = r - z_1$ and $ |z_2 + r| = z_2 + r$ in this region of integration. Moreover, performing a Taylor expansion in $z_{\pm}$, we get
\begin{equation}
\dfrac{1}{\sqrt{z_{\pm} \pm r}} \approx \dfrac{1}{\sqrt{r}} \pm \dfrac{z_{\pm}}{2 r^{3/2}}.
\end{equation}
The second term of this expansion can be neglected as $r$ is large. We write then
\begin{align}
  &\begin{aligned}
& \braket{\chi | 2 \kappa u_1 u_2} \approx \\
& 2 \kappa \dfrac{e^{-\beta r}}{\sqrt{r}} \left[ \int_{-r/2}^{r/2} \chi^{-}(z_1) u_1(z_1) e^{\beta z_1} \mathrm{d} z_1 +  \int_{-r/2}^{r/2} \chi^{+}(z_2)  e^{-\beta z_2}  u_2(z_2) \mathrm{d} z_2 \right],\\
  \end{aligned}
\end{align}
i.e., $\bra{\chi}\ket{2 \kappa u_1 u_2} \approx 2 \kappa \dfrac{e^{-\beta r}}{\sqrt{r}} I_1$, with $I_1$ defined as in the previous section. Proceeding in the same way with the other integrals from $\braket{\chi | b}=0$, we get
\begin{equation}
\dot{r} = A \dfrac{e^{- \beta r}}{\sqrt{r}} + B \dfrac{e^{-2 \beta r}}{r},
\label{dr-2D}
\end{equation}
with $A$ and $B$ as in the one dimensional case.

\end{document}